\documentclass{rmf-d}
\usepackage{nopageno,rmfbib,multicol,times,epsf,amsmath,amssymb,cite}
\usepackage{color}
\usepackage{xcolor}
\usepackage[latin1]{inputenc}
\usepackage[]{caption2}
\usepackage{graphics}
\usepackage{float}
\usepackage{graphicx}
\DeclareMathAlphabet{\mathpzc}{OT1}{pzc}{m}{it}
\def\rmfcornisa{Research OR Education of Physics \hfill\rmf\ 
\hfill MES A\~NO}

\def\rmfcintilla{{\it Rev.\ Mex.\ Fis.\ S.\/}}
\clearpage \rmfcaptionstyle 
\begin{document}
\title{ Dyson-Schwinger equations and the muon g-2
\vspace{-6pt}}
\author{ Kh\'epani Raya  }
\address{ Departamento de F\'isica Te\'orica y del Cosmos, Universidad de Granada, E-18071, Granada, Spain}
\author{ Adnan Bashir}
\address{ Instituto de F\'isica y Matem\'aticas, Universidad Michoacana de San Nicol\'as de Hidalgo, Morelia, Michoac\'an 58040, Mexico }
\author{ \'Angel S. Miramontes}
\address{ Instituto de F\'isica y Matem\'aticas, Universidad Michoacana de San Nicol\'as de Hidalgo, Morelia, Michoac\'an 58040, Mexico }
\author{ Pablo Roig}
\address{Departamento de F\'isica, Centro de Investigaci\'on y de Estudios Avanzados del IPN,\\ Apdo. Postal 14-740,07000 Ciudad de M\'exico, Mexico }

\maketitle
\recibido{XX YY 2021}{ZZ WW 2022
\vspace{-12pt}}
\begin{abstract}
\vspace{1em} We present a brief introduction to the Dyson-Schwinger equations (DSEs) approach to hadron and high-energy physics. In particular, how this formalism is applied to calculate the electromagnetic form factors $\gamma^* \gamma^* \to \textbf{P}^0$ and $\gamma^* \textbf{P}^\pm \to \textbf{P}^\pm$ (with $\textbf{P}^\pm$ and $\textbf{P}^0$ charged and neutral  ground-state pseudoscalar mesons, respectively) is discussed. Subsequently, the corresponding contributions of those form factors to the muon anomalous magnetic moment ($g-2$) are estimated. We look forward to promoting the DSE approach to address theoretical aspects of the muon $g-2$, highlighting some calculations that could be carried out in the future. \vspace{1em}
\end{abstract}
\keys{Dyson-Schwinger equations, electromagnetic form factors, hadronic light-by-light contributions \vspace{-4pt}}
\pacs{   \bf{\textit{13.40.Em Electric and magnetic moments; 
13.40.Gp Electromagnetic form factors; 11.10.St Bound and unstable states; Bethe-Salpeter equations}}    \vspace{-4pt}}
\begin{multicols}{2}

\section{Introduction}
The Standard Model (SM) of particle physics is supposed to describe ordinary matter in terms of elementary particles and their interactions. As a theory, the SM has been quite successful: continuous confrontation with empirical observations (experiment) reveals how robust it is. Even though it is sometimes difficult for it to explain some facts of Nature, such as confinement and dynamical mass generation in quantum chromodynamics (QCD)~\cite{Roberts:2021xnz}, only those quantities which require high precision measurements (or calculations) give us a small window to doubt the completeness of the SM to describe ordinary matter phenomena (aside from e.g. the origin of neutrino masses or the baryogenesis mechanism, which can be related to scales not directly accessible by current experiments). This is the case of the so called muon anomalous magnetic moment, $a_\mu:=(g_\mu-2)/2$.  The historical discrepancy between experimental and theoretical values of $a_\mu$ is fueled after the recent (first) measurement from the muon $g-2$ experiment at Fermilab (FNAL)~\cite{Muong-2:2021ojo}:
\begin{equation}
   a_\mu^{\mathrm{FNAL}}=116592040(54)\times10^{-11}\,,
\end{equation}
which, in combination with the previous experiment at Brookhaven National Laboratory (BNL)~\cite{Muong-2:2006rrc}, yields the accepted worldwide experimental average of:
\begin{equation}
 a_\mu^{\mathrm{Exp}}=116592061(41)\times10^{-11}\;.
\end{equation}
Conversely, several theoretical efforts, orchestrated by the Muon g-2 Theory Initiative, produce the following value as the Standard Model prediction of this quantity~\cite{Aoyama:2020ynm}:
\begin{equation}
\label{eq:SMprediction}
    a_\mu^{\mathrm{SM}}=116591810(43)\times10^{-11}\,.
\end{equation}
Experiment so far has reached an amazing precision of $0.35$ parts per million, and 
both SM and experiment exhibit the same level of accuracy; nevertheless, $a_\mu^{\mathrm{SM}}$ deviates $4.2\,\sigma$ from $a_\mu^{\mathrm{exp}}$. While waiting for ongoing runs at FNAL and upcoming new experiments at J-PARC~\cite{Saito:2012zz}, one should inquire on this observable from the theoretical point of view. The degree of sophistication of the SM experiment and predictions (about $a_\mu$), and more than anything, their discrepancy, excites those seeking explanations based on new physics. Either way, one must ensure that all contributions to $a_\mu$ that come from the SM are 
included and carefully calculated, as well as that they have a sufficient degree of precision. The SM contributions to $a_\mu$ are divided into those coming from quantum electrodynamics (QED), electroweak (EW) and hadronic contributions (at a fundamental level, QCD); the latter, further divided in hadronic vacuum polarization (HVP) and hadronic light-by-light (HLbL) contributions. Naturally, the QED part (calculated up to 5-loops) dominates~\cite{aoyama:2012wk}, $a_\mu^{\mathrm{QED}}=116 584 718.931(104)\times 10^{-11}\;.$ The EW contribution is subdominant but also well determined, $a_\mu^{\mathrm{EW}}=153.6(1.0)\times 10^{-11}$~\cite{czarnecki:2002nt,gnendiger:2013pva}. On the other hand, the hadronic contributions saturate the error in the SM prediction (see~\cite{Aoyama:2019ryr,Aoyama:2020ynm} and references therein):
\begin{eqnarray}
&a_\mu^{\mathrm{HVP}}=6845(40)\times 10^{-11}\;,&\\
    &a_\mu^{\mathrm{HLbL}}=92(18)\times 10^{-11}\;.&
\end{eqnarray}
It is not surprising that these are the least restricted contributions, demanding special attention. The non-perturbative nature of QCD often leads to difficulties not present in the QED-EW part of the SM. Thus, in evaluating the hadronic contributions to $a_\mu$ it is typical to appeal to data-driven analyses, effective theories, and lattice QCD; \emph{e.g.}~\cite{Davier:2019can,Keshavarzi:2019abf,Masjuan:2017tvw,Roig:2019reh,Guevara:2018rhj,Roig:2014uja, Danilkin:2019mhd,Colangelo:2017fiz,Masjuan:2020jsf,Miranda:2020wdg}. Nevertheless, there is a mathematical framework that requires neither the experimental inputs of the data-driven approaches nor the computer power of lattice QCD: the Dyson-Schwinger equations (DSEs) formalism~\cite{Roberts:1994dr,Fischer:2018sdj,Sanchis-Alepuz:2017jjd}. Even though there is a plethora of hadron physics predictions based upon this formalism (for example~\cite{Raya:2021zrz,Cui:2020tdf,Raya:2019vwr,Ding:2019qlr,Qin:2019hgk,Eichmann:2016yit,Raya:2015gva,Chang:2013nia,Chang:2013pq,Eichmann:2011vu}), as well as recent and pioneering works on hadronic contributions to the muon g-2~\cite{Miramontes:2021exi,Raya:2019dnh,Eichmann:2019tjk,Eichmann:2019bqf,Eichmann:2014ooa,Goecke:2010if}, little attention has been given to this approach as a valuable tool for muon $g-2$ related studies.

In this manuscript, we revisit Refs.~\cite{Raya:2019dnh,Miramontes:2021exi}, to briefly describe how to evaluate the HLbL $\textbf{P}^\pm$-\emph{box} contributions ($\textbf{P}^\pm=\pi^\pm,\,K^\pm$), given by the $\gamma^* \textbf{P}^\pm \to \textbf{P}^\pm $  elastic form factors (EFFs), and the $\textbf{P}^0$-\emph{pole} contributions ($\textbf{P}^0=\pi^0, \eta, \eta', \eta_c, \eta_b$), obtained from $\gamma^* \gamma^* \to \textbf{P}^0$  transition form factors (TFFs). The manuscript is organized as follows: Section 2 presents some general aspects for the description of mesons within the DSE framework. In Section 3, we discuss the calculation of EFFs and TFFs, presenting the obtained numerical results. Section 4 gathers such results and evaluates their contribution to $a_\mu$. Finally, Section 5 summarizes our findings and presents a vision about how the DSEs formalism can assist some theoretical aspects of the muon $g-2$.

\section{The Dyson-Schwinger and Bethe-Salpeter equations approach}
Essentially, the DSEs are the equations of motion in a quantum field theory; in this case, QCD. Every Green Function obeys a DSE which, in turn, requires the knowledge of at least one higher-order Green Function~\cite{Roberts:1994dr}. This forms an infinite tower of coupled (integral) equations which contains all the dynamics, therefore requiring a systematic truncation in order to extract the encoded physics~\cite{Qin:2020rad,Binosi:2016rxz}. Notwithstanding, this formalism captures the perturbative and non-perturbative facets of QCD at once, thus being an ideal platform to investigate hadron properties.

That being said, let us start by recalling the DSE for the quark propagator:
\begin{eqnarray}
    S^{-1}(p)&=&Z_2 [S^{(0)}(p)]^{-1} + \int_{q}^\Lambda K^{(1)}(q,p) S(q)\;, \nonumber \\\label{eq:quarkPropQCD}
    K^{(1)}(q,p)&=&\frac{4}{3} Z_1 g^2 D_{\mu\nu}(p-q) \gamma_\mu\otimes  \Gamma_\nu(p,q)\;,
\end{eqnarray}
where $\int_{q}^\Lambda = \int^\Lambda \frac{d^4q}{(2\pi)^4}$ stands  for  a  Poincar\'e  invariant  regularized  integration,  with  $\Lambda$  for the regularization scale. The rest of the pieces carry their usual meanings (color and flavor indices have been omitted for simplicity):
\begin{itemize}
    \item $D_{\mu\nu}$ is the gluon propagator and $g$ is the Lagrangian coupling constant.
    \item $\Gamma_\nu$ the fully-dressed quark-gluon vertex (QGV) which, in its full glory, 
    is characterized by 12 Dirac structures~\cite{Albino:2021rvj,AtifSultan:2018end,  Albino:2018ncl} (some of them are explicitly connected with dynamical mass generation in QCD).
    \item $Z_{1,2}$ are the QGV and quark wave-function renormalization constants, respectively.
\end{itemize}
Herein, $S^{(0)}$ denotes the bare quark propagator,
\begin{equation}
    S^{(0)}(p)=[i \gamma \cdot p+m^{\text{bm}}]^{-1}\;,
\end{equation}
where $m^{\text{bm}}$ is the Lagrangian bare mass. The fully-dressed quark propagator is represented as
\begin{equation}
    \label{eq:quarkPropDef}
    S(p)=  Z(p^2) (i \gamma \cdot p + M(p^2))^{-1}\;,
\end{equation}
in such a way that the non-perturbative effects of the strong interactions are captured in the dressing functions  $Z(p^2)$ and $M(p^2)$, in analogy with its bare counterpart. In fact, the mass function, $M(p^2)$, is enhanced (a couple hundred MeVs) in the infrarred region, as a consequence of dynamical chiral symmetry breaking (DCSB)~\cite{Roberts:2021xnz}. From all the functions appearing in Eq.~\eqref{eq:quarkPropQCD}, only the quark mass function is independent of the renormalization point $\zeta$. 

 The description of mesons is obtained from the Bethe-Salpeter equation (BSE)~\cite{Eichmann:2016yit,Qin:2020rad}:
\begin{equation}
\label{eq:BSEGen}
    \Gamma_{H}(p;P)=\int_{q}^\Lambda K^{(2)}(q,p;P)\chi_{H}(q;P)\;,
\end{equation}
whose ingredients are defined as follows:
\begin{itemize}
    \item $\Gamma_{H}$ corresponds to the  Bethe-Salpeter amplitude (BSA), $H$ labeling the type of meson, while $\chi_{H}(q;P)=S(q_+)\Gamma_{H}(q;P)S(q_-)$ denotes the BS wavefunction (BSWF).
    \item $P$ is the total momentum of the  bound state; $q_+:=q+\eta P$ and $q_-:=q-(1-\eta)P$, with $\eta\in[0,1]$ defining the relative momentum. 
\end{itemize}
The Dirac structure characterizing the BSA depends on the meson's quantum numbers. In particular, for a pseudoscalar meson $\textbf{P}$:
    \begin{eqnarray}
\label{eq:BSA}
    \Gamma_{\textbf{P}}(q;P) &=& \gamma_5[i E_{\textbf{P}}(q;P)+\gamma \cdot P F_{\textbf{P}}(q;P) \\
    &+& \gamma \cdot q G_{\textbf{P}}(q;P) + q_\mu \sigma_{\mu \nu} P_\nu H_{\textbf{P}}(q;P)]\;,\nonumber
\end{eqnarray}
where that amplitude attached to $\gamma_5$, $E_\textbf{P}(q;P)$, is dominant. Finally, $K^{(2)}$ corresponds to two particle irreducible quark/antiquark scattering kernel, which expresses the interactions between the quark and antiquark within the bound-state. Both Eq.~\eqref{eq:quarkPropQCD} and Eq.~\eqref{eq:BSEGen} require, in principle, the knowledge of infinitely many QCD's Green Functions. A truncation scheme must be then specified in order to arrive at a tractable problem. Symmetry principles establish that, in fact, $K^{(1)}$ and $K^{(2)}$ are connected via vector and axial-vector Ward-Green-Takahashi identities (WGTIs)~\cite{Xing:2021dwe,Qin:2014vya, Bhagwat:2007ha}; the vector WGTI entails electric charge conservation, while the axial-vector one implies the appearance of pions (light pseudoscalars) as  the Goldstone bosons of DCSB~\cite{Bender:1996bb}. 

Once a systematic truncation is chosen, the meson mass and BSWF are obtained from solutions of Eqs.~(\ref{eq:quarkPropQCD}) and (\ref{eq:BSEGen}); furthermore, the pseudoscalar meson leptonic decay constant ($f_\textbf{P}$) can be computed straightforwardly from the canonically normalized BSWF as:
\begin{equation}
    f_{\textbf{P}}P_{\textbf{P}}= \text{tr}_{CD} Z_2 \int_q \gamma_5 \gamma_\mu \chi_{\textbf{P}}(q;P)\;,
\end{equation}
where $\text{tr}$ indicates the trace over color and Dirac indices. In the next section we briefly explain some sensible truncations for ground-state pseudoscalar mesons, employed in subsequent calculations.

\subsection{Rainbow Ladder truncation}
The simplest truncation that fullfils vector and axial-vector WGTIs is defined by the kernel ( $\{t,\,u,\,r,\,s\}$ color indices):
\begin{eqnarray}
\label{eq:defRL}
    [K_{tu}^{rs}]^{\text{RL}}(q,p;P)=-\frac{4}{3} Z_2^2 D_{\mu\nu}^{\text{eff}}(p-q)[\gamma_\mu]_{ts} \otimes [\gamma_\nu]_{ru} \;,
\end{eqnarray}
which relate the 1-body and 2-body kernels as:
\begin{equation}
\label{eq:defRL2}
    K^{(2)}(q,p;P)=K^{\text{RL}}(q,p;P)=-K^{(1)}(q,p;P)\;.
\end{equation}
This truncation is dubbed as the RL truncation~\cite{Bender:1996bb}; a sensible and practical approach so long as we restrain ourselves to ground-state pseudoscalar and vector mesons~\cite{Xu:2019ilh,Ding:2019qlr,Raya:2015gva,Chang:2013nia}. It is worth noticing that the gluon propagator has been promoted to an effective one, $g\,D_{\mu\nu} \to D_{\mu\nu}^{\text{eff}}$, where:
\begin{equation}
    D_{\mu\nu}^{\text{eff}}(k) =\left( \delta_{\mu \nu}-\frac{k_\mu k_\nu}{k^2} \right)\mathcal{G}(k^2)\;.
\end{equation}
Herein, $\mathcal{G}(k^2)$ captures all the missing information from the rich structure of $\Gamma_\nu$, lost after reducing $\Gamma_\nu \to \gamma_\nu$. Typically, we appeal to lattice QCD or phenomenological models to provide a sound representation for $\mathcal{G}(k^2)$~\cite{Serna:2018dwk,Chang:2021vvx,Qin:2011dd}. Throughout this work, we shall employ the so called Qin-Chang (QC) interaction~\cite{Qin:2011dd}. Avoinding details, the QC model is defined once the strength parameter, $m_G = (w D)^{1/3}$, is fixed to produce the masses and decay constants of the ground-state pseudoscalar mesons. Typical RL parameters are $m_G \sim 0.8$ GeV and $w \sim 0.5$ GeV; herein, the later is varied within the range $w\in(0.4,\,0.6)$ to estimate model uncertainties.

\subsection{Beyond RL: Anomaly kernel}
In discussing the $\eta-\eta'$ states, it is convenient to work with a flavor basis, such that the associated Bethe-Salpeter amplitudes can be expressed as ($l=u=d$):
	\begin{eqnarray}
	\Gamma_{\eta,\eta'}(k;P) &=& \textrm{diag}(1,1,0) \Gamma_{\eta,\eta'}^l(k;P)\\
	&+&\textrm{diag}(0,0,\sqrt{2}) \Gamma_{\eta,\eta'}^s(k;P)\;.
	\end{eqnarray}
Following Refs.~\cite{Bhagwat:2007ha,Ding:2018xwy}, the effects of the non-Abelian anomaly, not present in $K^{\text{RL}}$ alone, are introduced at the level of the  Bethe-Salpeter kernel ($k = p-q$):
\begin{eqnarray} \label{eq:anomaly}
[K_{tu}^{rs}]^A(q,p;P) &=& -\mathcal{G}_A(k^2)\huge( \sin^2{\theta_\xi} [\mathbf{r} \gamma_5]_{rs}  [\mathbf{r} \gamma_5]_{tu} \\
	&+& \frac{1}{\chi_l^2} \cos^2{\theta_\xi} [\mathbf{r} \gamma_5 \gamma \cdot P]_{rs}  [\mathbf{r} \gamma_5 \gamma \cdot P]_{tu} \huge)\;, \quad \nonumber
\end{eqnarray}
where $\chi_l = M_l(0)$, and $\theta_\xi$ controls the relative strength between the $\gamma_5$ and $\gamma_5 \gamma \cdot P$ terms; $\mathbf{r}=$diag$(1,1,\nu_R)$ and $\nu_R = M_l(0)/M_s(0) = 0.57$. The strength and momentum dependence of the anomaly is controlled by
	\begin{equation}
	\mathcal{G}_A(k^2) = \frac{8 \pi^2}{\omega_\xi^4} D_\xi\; \textrm{exp}[-k^2/\omega_\xi^2]\;.
	\label{eq:anomaly2}
	\end{equation}
Thus, the 2-body kernel for the $\eta-\eta'$ case becomes:
\begin{equation}\label{eq:anomaly2}
    K^{(2)}(q,p;P) = K^{\text{RL}}(q,p;P)+K^A(q,p;P)\;.
\end{equation}
The new parameters,  $\{D_\xi,\;\omega_\xi,\;\cos^2 \theta_\xi\}$, are fixed to provide a fair description of $m_{\eta,\eta'}$ and $f_{\eta,\eta'}^{l,s}$~\cite{Raya:2019dnh,Ding:2018xwy}.

\subsection{Beyond RL: Meson cloud effects}
The interaction of a photon with a quark, the quark-photon vertex (QPV), is described by an inhomogeneous BSE:
\begin{equation}
\label{eq:BSEin}
    \Gamma_{\mu}^f(p;P)=\gamma_\mu+\int_{q}^\Lambda K^{(2)}(q,p;P)\chi_\mu^f(q;P)\;,
\end{equation}
where $\Gamma_\mu^f$ represents the QPV, the interaction of a photon with a $f$-flavor quark; $\chi_\mu^f(q;P)$ is simply the unamputated vertex, which reads:
\begin{equation}
    \chi_\mu^f(q;P)=S^f(q_+)\Gamma_{\mu}^f(q;P)S^f(q_-)\;.
\end{equation}
In the RL truncation, bound-states obtained from the homogeneous BSE, Eq.~\eqref{eq:BSEGen}, appear as poles in the time-like axis. By corollary, solutions of Eq.\eqref{eq:BSEin} exhibit poles at $Q^2=-m_n^2$~\cite{Maris:1999bh} (herein, $m_n$ corresponds to vector meson masses, such that $m_1 = m_\rho$). This is an appreciated feature for space-like form factors; in particular, charge radii are obtained with better accuracy~\cite{Maris:1999bh,Maris:2000sk}. In the case of the $\gamma^* \gamma^*$ TFFs, Eq.~\eqref{eq:BSEin} also guarantees that the abelian anomaly is faithfully reproduced~\cite{Eichmann:2019tjk,Maris:2002mz}. This makes the RL truncation of QCD's DSE a sound treatment to estimate the $\textbf{P}^\pm$-\emph{box} and $\textbf{P}^0$-\emph{pole} contributions: only a relatively small space-like region of the corresponding form factors actually matters for determining their contribution to $a_\mu$~\cite{Eichmann:2019bqf,Raya:2019dnh}. 

It is worth exploring, however, what information a proper treatment of the time-like region could provide. The truncation introduced in Refs.~\cite{Miramontes:2019mco,Miramontes:2021xgn}, denoted herein as BRL, takes into account resonance effects, thus shifting  the vector-meson poles  to the complex plane; subsequently, meson cloud effects (MCEs) are incorporated in the description of the pion EFF~\footnote{MCEs are also expected to play a major role in the description of nucleon and hyperon transition form factors, around $Q^2\approx 0$~\cite{Williams:1993ux,Ramalho:2016zgc,Granados:2017cib,Raya:2021pyr}.}. These explorations have been adapted in~\cite{Miramontes:2021exi} to compute the $\pi-K$ EFFs and corresponding box contributions; in this case, the QC interaction model demands $m_G\sim 0.84$ and $\omega \in (0.6,\,0.8)$ GeV to accurately produce $m_{\pi,\,K}$ and $f_{\pi,\,K}$.

\section{Electromagnetic form factors}
Let us now focus on the calculation of the electromagnetic elastic and transition form factors, $\gamma^* \textbf{P}^\pm \to \textbf{P}^\pm$ and $\gamma^* \gamma^* \to \textbf{P}^0$, respectively.

\subsection{Elastic form factors}
The electromagnetic process $\gamma^* \textbf{P}^\pm \to \textbf{P}^\pm$ is described by a single form factor, $F_{\textbf{P}^\pm}(Q^2)$. In the impulse approximation, which corresponds to a triangle diagram and is self consistent with the RL truncation~\cite{Maris:2000sk}, $F_{\textbf{P}^\pm}(Q^2)$ is obtained from~\cite{Chang:2013nia}: 
\begin{eqnarray}
    \label{eq:defEFF}
   & 2K_\mu F_{\textbf{P}^\pm}(Q^2) = e_u [F_{\textbf{P}^\pm}^{u}(Q^2)]_\mu + e_{\bar{h}} [F_{\textbf{P}^\pm}^{\bar{h}}(Q^2)]_\mu\;,
\end{eqnarray}
where $\textbf{P}^\pm$ is a $u \bar{h}$ meson and $e_{u,\bar{h}}$ are the electric charges of the quark and antiquark, respectively. $[F_{\textbf{P}^\pm}^f(Q^2)]_\mu$ denotes the interaction of the photon with a valence constituent $f$-in-$\textbf{P}^\pm$, such that:
\begin{eqnarray}
    [F_{\textbf{P}^\pm}^f(Q^2)]_\mu =\nonumber \text{tr}_{CD} \int_q \chi_{\mu}^f(q+p_f,q+p_i)\\
    \times \Gamma_{\textbf{P}^\pm}(q_i;p_i)S(q)\Gamma_{\textbf{P}^\pm}(q_f;-p_f)\;. \label{eq:defEFF2}
\end{eqnarray}
The kinematics is defined as follows: $p_{i,f}=K\mp Q/2$ and $q_{i,f}=q+p_{i,f}/2$, such that $p_{i,f}^2=-m_{\textbf{P}^\pm}^2$; naturally, $m_{\textbf{P}^\pm}$ is the mass of the pseudoscalar meson and $Q$ the photon momentum. Beyond RL, Eq.~\eqref{eq:defEFF2} might be supplemented by additional terms~\cite{Maris:2000sk}, and so is the case of  the BRL truncation. Thus, in principle, all the necessary ingredients for the computation of $F_{\textbf{P}^\pm}(Q^2)$ have been gathered.

For future references, and having made use of Eq.~\eqref{eq:BSEin} to define the QPV, we will name this approach the \emph{direct computation}. This usual procedure was employed in Refs.~\cite{Eichmann:2019tjk, Eichmann:2019bqf} for the calculation of $\pi-K$ EFFs and corresponding box contributions, albeit in the RL truncation with the Maris-Tandy (MT) model~\cite{Maris:1999nt}.

\subsection{Transition form factors}
The transition $\gamma^{*}\gamma^{*} \to \textbf{P}^0$ is also described by a single form factor, $G_{\textbf{P}^0}(Q_1^2,Q_2^2, Q_1\cdot Q_2)$. In the impulse approximation~\cite{Raya:2015gva}:
	\begin{eqnarray}
	\mathcal{T}_{\mu\nu}(Q_1,Q_2)&=& T_{\mu\nu}(Q_1,Q_2) + T_{\nu\mu}(Q_2,Q_1)\;,\\ \nonumber
	T_{\mu\nu}(Q_1,Q_2) &=& \frac{1}{4\pi^2 }\epsilon_{\mu \nu \alpha \beta} Q_{1\alpha} Q_{2\beta} G_{\textbf{P}^0}(Q_1^2,Q_2^2, Q_1\cdot Q_2) \\ \nonumber
	&=& \mathbf{e}_{\textbf{P}^0}^2 \; \text{tr}_{CD}\int_q i  \chi_{\mu}^f(q,q_1)\Gamma_{\textbf{P}^0}(q_1,q_2) \\
	&\times& S_f(q_2) i \Gamma_\nu^f(q_2,q)\;,\label{eq:defTFF}
	\end{eqnarray}
	where $Q_{1},\;Q_2$ are photon momenta, and the kinematics is set by: $(Q_1+Q_2)^2=P^2 = - m_{\textbf{P}^0}^2$, $q_1=q+Q_1$, $q_2=q-Q_2$;  finally,  $\mathbf{e}_{\textbf{P}^0}$ is a  factor associated with the electric charges of the valence quark/antiquark~\footnote{Eq.~\eqref{eq:defTFF} must be adapted to account for the flavour decomposition of the $\eta-\eta'$ systems~\cite{Ding:2018xwy}.}.	As for the EFFs, the calculation of TFFs requires the knowledge of the quark propagators, BSAs and QPV.  This exhibits how, within the DSE formalism, hadronic observables maintain a traceable connection with QCD's fundamental ingredients.

A direct computation can be performed at this stage. Nevertheless, technical reasons restrict the evaluation of the form factors to a limited domain of space-like momenta. For instance, the pion elastic and single of shell TFFs can only be obtained up to $Q^2\sim4$ GeV$^2$~\cite{Maris:2000sk,Maris:2002mz}, without appealing to sophisticated mathematical techniques for extrapolation~\cite{Eichmann:2017wil}. The domain in which a direct computation of the form factors is possible is sufficient to accurately estimate their corresponding contributions to $a_\mu$. However, we also present an alternative technique, based upon perturbation theory integral representations (PTIRs)~\cite{Raya:2015gva,Chang:2013nia,Chang:2013pq}, to evaluate the form factors at arbitrarily large momenta. Among other things, this allows us to take into account the asymptotic behavior of the form factors when proposing parametric representation for the numerical data, particularly relevant in the case of TFFs~\cite{Raya:2019dnh, Masjuan:2017tvw,Knecht:2001qf}.

\subsection{The PTIR approach}
A practical  perturbation theory integral representation for the quark propagators and BSAs was put forward in~\cite{Chang:2013pq,Chang:2013nia}, to calculate the pion distribution amplitude and space-like EFF. The general idea, which applies to all pseudoscalars~\cite{Raya:2015gva,Raya:2016yuj,Ding:2018xwy}, is to describe the quark propagators in terms of $j_m=2$ complex conjugate poles (CCPs), and express the BSAs, $\mathcal{A}_j(k;P)$, as follows:
	\begin{subequations}\label{BSAPTIR}
		\begin{eqnarray}
		\mathcal{A}_j(k;P)= \sum_{i=1}^{i_n} \int_{-1}^1 dw \rho_i^j(w) \frac{c_i^j \, (\Lambda^2_{i,j})^{\beta_i^j}}{(k^2+w k \cdot P + \Lambda^2_{i,j})^{\alpha_i^j}}\;.
		\end{eqnarray}
	\end{subequations}
The interpolation parameters $\{z_j, m_j \}$, $\{\alpha_i^j, \beta_i^j, \Lambda_{i,j}, c_i^j, i_n = 3 \}$ (for quark propagators and  BSAs, respectively), as well as the spectral weights, $\rho_i^j(w)$, are determined through fitting to the numerical results of the corresponding DSE-BSEs. The sets of parameters are found through Refs.~\cite{Raya:2019dnh,Miramontes:2021exi,Shi:2014uwa}.

\begin{figure}[H]
 \centering
 \includegraphics[width=\linewidth]{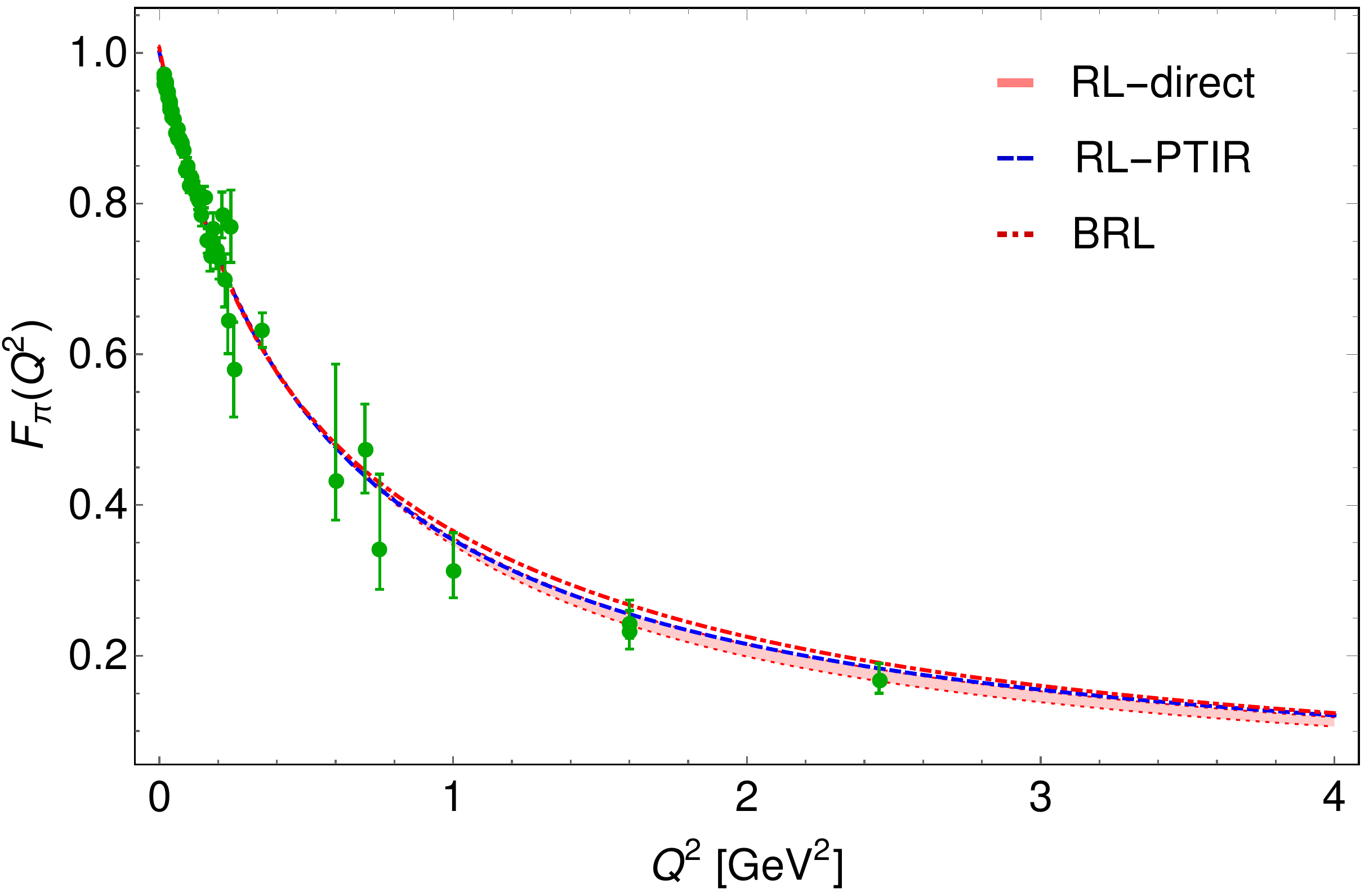}\\
 \includegraphics[width=\linewidth]{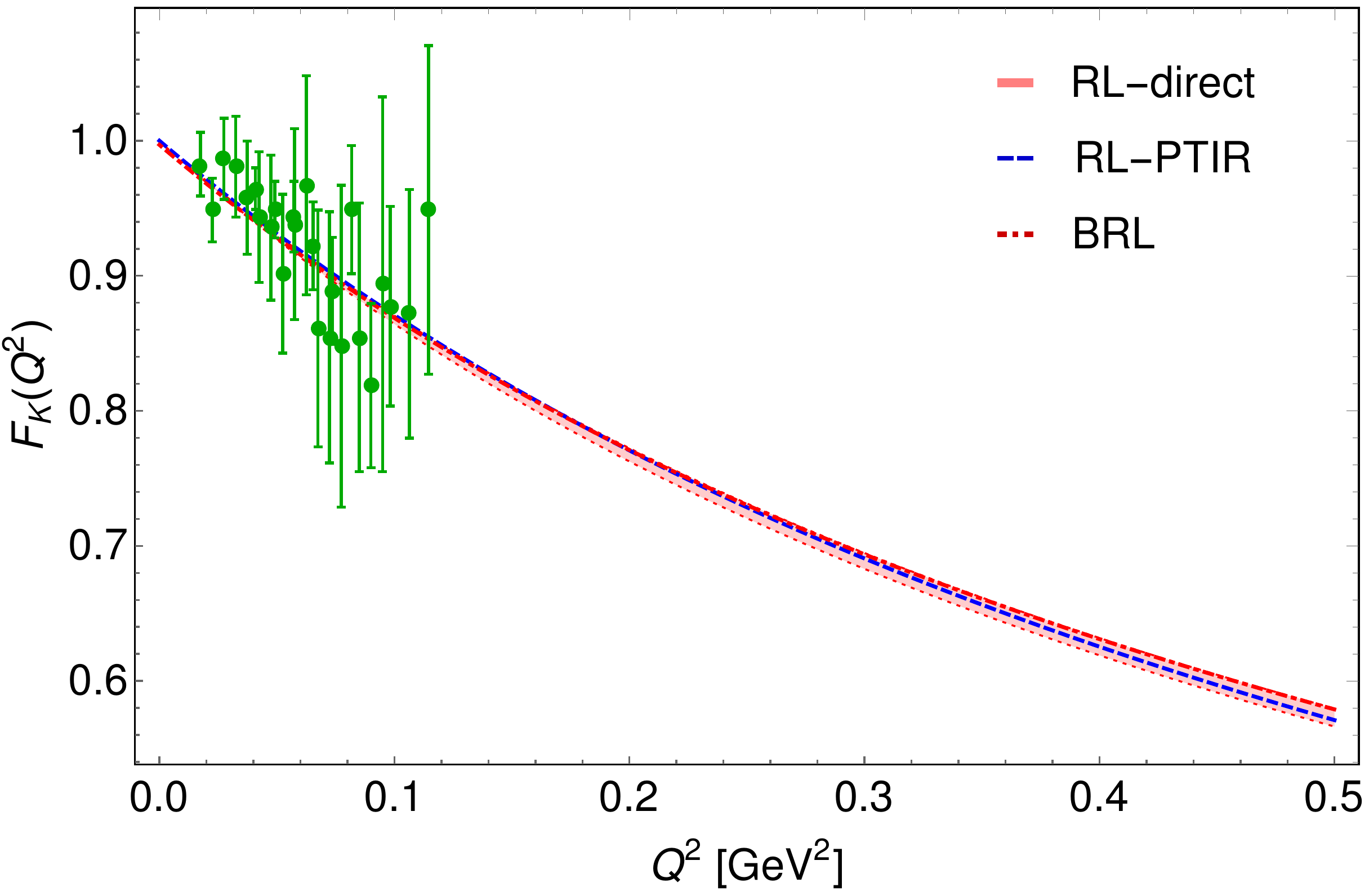}\\
 \caption{ $\pi^+$ and $K^+$ EFFs. The band in the RL-direct result accounts for the variation of the QC model parameters, as described in text; those corresponding to the PTIR and BRL results are not shown, since there is a considerable overlap. The charge radii, $r_\pi = 0.676(5)$ fm and $r_K= 0.595(5)$ fm are practically insensitive to the model inputs and truncation. Experimental data from Refs.~\cite{Dally:1980dj,Amendolia:1986ui,NA7:1986vav,JeffersonLab:2008gyl}.}
 \label{fig:EFFs}
\end{figure}

	\begin{figure}[H]
		\centering
		\includegraphics[width=0.5\textwidth]{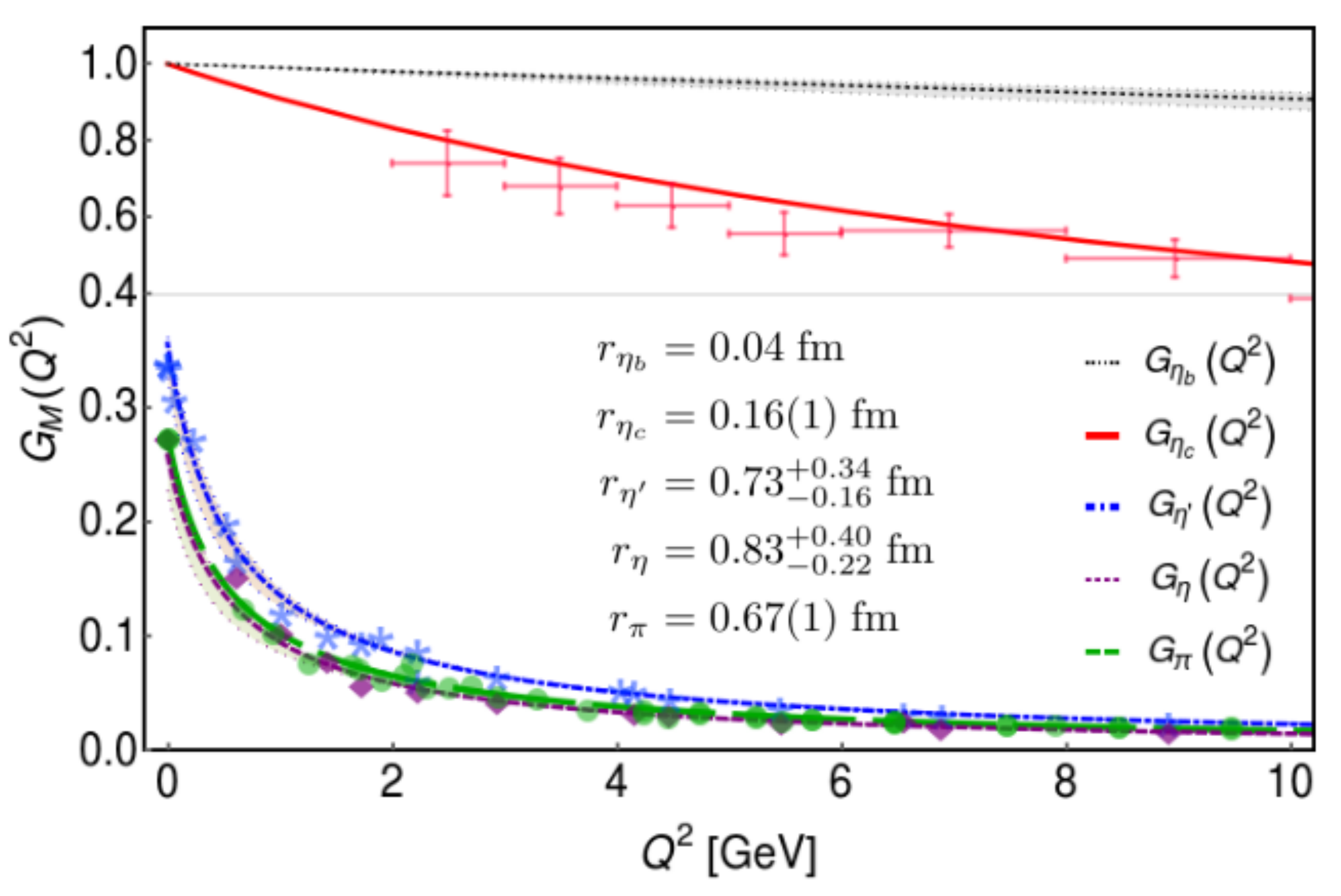}
		\caption{Single off-shell $\gamma\gamma^* \to \textbf{P}^0$ TFFs and corresponding charge radii~\cite{Raya:2019vwr}. Experimental data is taken from~\cite{Danilkin:2019mhd} and references therein.  The mass units are in GeV.}
		\label{fig:TFFs}
		\centering
	\end{figure}

Constructing a PTIR for the QPV in Eq.~\eqref{eq:BSEin} turns out to be difficult and unpractical~\cite{Xu:2019ilh}. Thus, appealing to gauge covariance properties~\cite{Delbourgo:1977jc}, the following Ansatz has been proposed and systematically tested~\cite{Raya:2015gva,Raya:2016yuj,Ding:2018xwy}:
	\begin{eqnarray}
	\nonumber
	\chi_\mu(k_f,k_i)&=& \gamma_\mu \Delta_{k^2 \sigma_V} \\
	\nonumber
	&+& [\mathbf{s} \gamma\cdot k_f \gamma_\mu \gamma \cdot k_i + \bar{\mathbf{s}}\gamma\cdot k_i \gamma_\mu \gamma \cdot k_f]\Delta_{\sigma_V}\nonumber\\
	\nonumber
	&+&[\mathbf{s}(\gamma\cdot k_f \gamma_\mu + \gamma_\mu \gamma \cdot k_i)\\
	&+&\bar{\mathbf{s}}(\gamma\cdot k_i \gamma_\mu + \gamma_\mu \gamma \cdot k_f)]i\Delta_{\sigma_S}\;,
	\label{eq:vertex}
	\end{eqnarray}
	where $\Delta_F=[F(k_f^2)-F(k_i^2)]/(k_f^2-k_i^2)$, $\bar{\mathbf{s}}=1-\mathbf{s}$. Up to transverse pieces associated with $\mathbf{s}$, $\chi_\mu(k_f,k_i)$ and $S(k_f)\Gamma_\mu(k_f,k_i)S(k_i)$ are equivalent. The transverse terms are weighted by the function $\mathbf{s}(Q_1^2,Q_2^2)$, which is exponentially suppresed and tuned to reproduce the empirical decay widths~\cite{Raya:2019dnh}; as a by-product, the effects of the $\rho$-meson pole in the low-$Q^2$ domain are properly mimicked.

Defined as in Eq.~\eqref{eq:vertex} - flavor labels omitted on purpose, the QPV is expressed in terms of the quark propagator dressing functions~\footnote{The quark propagator is  written as $S(p) = -  i \gamma \cdot p\;\sigma_v(p^{2}) + \sigma_s(p^{2})$, with $\sigma_{s,v}(p^2)$ being algebraically related to $M(p^2)$ and $Z(p^2)$ in Eq.~\eqref{eq:quarkPropDef}.}. With all the ingredients in Eqs.~(\ref{eq:defEFF}, \ref{eq:defEFF2}, \ref{eq:defTFF}) expressed in a PTIR, the evaluation of the 4-momentum integral follows after a series of standard algebraic steps (numerical integration is only carried out for the Feynman parameters and spectral weights). Hence, the form factors can be calculated at arbitrarily large space-like momenta.

\section{Numerical Results and HLbL contributions}

\subsection{Elastic and transition factors}
The $\pi$ and $K$ EFFs are presented in Fig.~\ref{fig:EFFs}. We compare the RL results which follow from the direct computation and PTIR approach; the compatibility between both calculations is evident. In the depicted domain, the BRL truncation yields similar outcomes. Furthermore, our obtained EFFs are in clear agreement with the DSE results reported in Ref.~\cite{Eichmann:2019bqf}. TFFs are presented in Fig.~\ref{fig:TFFs}, in the single off-shell case. The agreement of those form factors with the available experimental data is clear. Notably, the PTIR approach enabled  us to calculate the $\eta_c$ and $\eta_b$ TFFs without facing new obstacles.


\subsection{Charged pion and kaon pole contributions}
To calculate the $\textbf{P}^\pm$-box contributions, we employ the master formula derived in~\cite{Colangelo:2017fiz}, which reads:
\begin{equation}
\alpha_{\mu}^{\textbf{P}-box} = \frac{\alpha^3_{em}}{432 \pi^2} \int_{\Omega}  \sum_i^{12} T_i(Q_1,Q_2,\tau) \bar{\Pi}_i^{\textbf{P}-box} (Q_1,Q_2,\tau),
\label{eq.master_formula}
\end{equation}
where $\alpha_{em}$ is the QED coupling constant, and
\begin{equation}
    \int_{\Omega}:= \int_{0}^{\infty} d\tilde{\Sigma}~\tilde{\Sigma}^3 \int_0^1 dr~r\sqrt{1-r^2} \int_0^{2\pi}d\phi\;.
\end{equation}
The functions $\bar{\Pi}_i^{\text{P}-box}$ are expressed as:
\begin{eqnarray}\nonumber
\bar{\Pi}_i^{\textbf{P}-box}(Q_1^2,Q_2^2,Q_3^2) = F_{\textbf{P}}(Q_1^2)F_{\textbf{P}}(Q_2^2) F_{\textbf{P}}(Q_3^2)&\\
\times \frac{1}{16 \pi^2} \int_0^1 dx \int_0^{1-x} dy I_i(x,y)\,.&
\label{I_feynman}
\end{eqnarray}
The scalar functions $T_i$ and $I_i$ are provided in Appendices B and C of  Ref.~\cite{Colangelo:2017fiz}, respectively. With the EFFs obtained in the RL truncation (direct and PTIR) and BRL, the numerical estimates for the $\pi^\pm-$box contributions are:
\begin{eqnarray} \nonumber
a_\mu^{\pi^{\pm}-\text{box}} &=& -(15.4\pm 0.3) \times 10^{-11} \;\;\text{[RL-direct]}\;, \\
a_\mu^{\pi^{\pm}-\text{box}} &=& -(15.6\pm 0.3) \times 10^{-11} \nonumber \;\;\text{[RL-PTIR]}\;, \\
a_\mu^{\pi^{\pm}-\text{box}} &=& -(15.7\pm 0.2) \times 10^{-11} \;\;\text{[BRL]}\;. \label{eq:boxpi}
\end{eqnarray}
Analogous results for the $K^\pm$ case yield:
\begin{eqnarray}
a_\mu^{K^{\pm}-\text{box}} &=& -(0.47\pm 0.03) \times 10^{-11} \nonumber \;\;\text{[RL-direct]}\;, \\
a_\mu^{K^{\pm}-\text{box}} &=& -(0.48\pm 0.03) \times 10^{-11} \nonumber \;\;\text{[RL-PTIR]}\;, \\
a_\mu^{K^{\pm}-\text{box}} &=& -(0.48 \pm 0.02) \times 10^{-11} \;\;\text{[BRL]}\;.\label{eq:boxkaon}
\end{eqnarray}
From Fig.~\ref{fig:EFFs} and the above estimates, it is clear that the direct and PTIR approach are plainly compatible; the BRL truncation also yields similar outcomes.  Therefore, one can combine the estimates in Eqs.~\eqref{eq:boxpi}-\eqref{eq:boxkaon} to produce the weighted averages:
\begin{eqnarray}
    a_\mu^{\pi^\pm-\text{box}}&=&-(15.6\pm 0.2)\times 10^{-11}\;,\\
    a_\mu^{K^\pm-\text{box}}&=&-(0.48\pm 0.02)\times 10^{-11}\;,
\end{eqnarray}
where the errors reflect  model and truncation uncertainties.

\subsection{Pseudoscalar pole contributions}
The master formula for the $\textbf{P}^0$-pole contributions is found in Refs.~\cite{Knecht:2001qf,Roig:2014uja}. It is highly convenient to parameterize the numerical results of the TFFs in a sensible way. For the light pseudoscalars, $\{\pi^0, \eta, \eta' \}$, a  Canterbury approximants representation turns out to be quite adequate~\cite{Masjuan:2017tvw}; with proper care, it captures all the short- and long distance facets of the TFFs~\cite{Raya:2019dnh}. Those form factors are then represented as 
($x=Q_1^2$, $y=Q_2^2$):
	\begin{eqnarray}\nonumber
	G_\text{P}^{\text{light}}(x,y) = \frac{a_{00}+a_{10}(x+y)+a_{01}(xy)}{1+b_{10}(x+y)+b_{01}(xy)+b_{11}(x+y)(xy)}\;, \label{eq:CAsdef}
	\end{eqnarray}
where the independent parameters, $b_{i,j}$, are fitted to the numerical data ($a_{i,j}$ are fixed by the axial anomaly and  short-distance QCD constraints, properly captured by the actual numerical solutions). The hardness of the $\{ \eta_c,\;\eta_b\}$ TFFs demands a much simpler representation. All interpolation parameters are listed in~\cite{Raya:2019dnh}. Finally, the corresponding $\textbf{P}^0$-\emph{pole} contributions are:
	\begin{eqnarray}
	\nonumber
	a_\mu^{\pi^0-\textrm{pole}}&=&\left(61.4\pm0.21 \right)\times 10^{-11}\;,\\
	\nonumber
	a_{\mu}^{\eta-\textrm{pole}} &=& (14.7 \pm 0.19) \times 10^{-11}\;, \\
	\nonumber
	a_{\mu}^{\eta'-\textrm{pole}} &=& (13.6 \pm 0.08) \times 10^{-11}\;, \\
	\nonumber
	a_{\mu}^{\eta_c-\textrm{pole}} &=& (0.9 \pm 0.1) \times 10^{-11}\;, \\
	a_{\mu}^{\eta_b-\textrm{pole}} &=& (0.26 \pm 0.01) \times 10^{-13}\;. \label{eq:neutralcont}
	\end{eqnarray}
Summing up the results from~\eqref{eq:neutralcont}:
\begin{equation}
    a_\mu^{\textbf{P}-\text{pole}}=(90.6\pm4.9)\times 10^{-11}\;,
\end{equation}
where the errors, accounting for model uncertainties (mostly dominated by the QPV Ansatz), have been added linearly.

\section{Conclusions and scope}
We described the computation of the electromagnetic form factors $\gamma^* \textbf{P}^\pm \to \textbf{P}^\pm$ and $\gamma^* \gamma^* \to \textbf{P}^0$, within the DSE approach to QCD, aiming to evaluate their contributions to $a_\mu$. 

The EFFs were obtained, firstly, in the RL truncation. Direct computations and the PTIR approach were shown to be fully compatible, while also being in agreement with the DSE results from Refs.~\cite{Eichmann:2019tjk,Eichmann:2019bqf}. It was also confirmed that the BRL truncation, which incorporates MCEs~\footnote{The MCEs take place in the neighborhood of $Q^2\approx 0$~\cite{Williams:1993ux,Ramalho:2016zgc,Granados:2017cib}, such that, for increasing $Q^2$, BRL $\to$ RL.}, produces   similar EFFs in the relevant domain for $a_\mu$; the value of the latter being barely affected by the new effects in the truncation. Our most recent analysis,~\cite{Miramontes:2021exi}, supports these observations.

The validity of the PTIR approach is also manifested in the case of the $\gamma^* \gamma^* \to \{\pi^0, \eta, \eta', \eta_c, \eta_b \}$ TFFs, whose calculation has been summarized herein and described in detail through Refs.~\cite{Raya:2015gva,Raya:2016yuj,Ding:2018xwy,Raya:2019dnh}. Our $\{\pi^0,\;\eta\;,\eta' \}$ computations are fully compatible with those from~\cite{Eichmann:2017wil,Eichmann:2019tjk}, even though the dealing with the non-Abelian anomaly is vastly different. Interestingly, our numerical results suggest a sizeable contribution from the $\eta_c$ meson, which might be worth exploring in the future.

In this manuscript we have briefly discussed the capabilities of the DSE formalism to address calculations of hadronic observables, highlighting some quantities of interest for the muon $g-2$. We hope to continue developing calculations related to the subject. For instance, following the remarks from~\cite{Roig:2019reh}, the contribution from axial-vector mesons is worth exploring, even though it represents a major computational challenge; on the other hand, estimating the contribution from excited pseudoscalars is within reach.

\section{Acknowledgements}
The authors acknowledge support from CONACYT and C\'atedras Marcos Moshinsky (Fundaci\'on Marcos Moshisnky). This research was also partly supported by Coordinaci\'on de la Investigac\'on Cientifica (CIC) of the University of Michoac\'an, Mexico, through Grant No. 4.10. KR wishes to acknowledge J. Rodr\'iguez-Quintero for his valuable scientific counselling. 

\end{multicols}
\medline
\begin{multicols}{2}
\bibliography{main}
\end{multicols}
\end{document}